\documentclass[conference]{IEEEtran}
\IEEEoverridecommandlockouts
\usepackage[hyphens]{url}
\usepackage[hidelinks]{hyperref}
\usepackage{adjustbox}
\hypersetup{breaklinks=true}
\usepackage{cite}
\usepackage{amsmath,amssymb,amsfonts}
\usepackage{graphicx}
\usepackage{subcaption}
\usepackage{textcomp}
\usepackage{xcolor}
\usepackage{url}
\usepackage{float}
\def\BibTeX{\rm B\kern-.05em{\sc i\kern-.025em b}\kern-.08em
    T\kern-.1667em\lower.7ex\hbox{E}\kern-.125emX}

\usepackage{algorithm}
\usepackage{algorithmicx}
\usepackage[noend]{algpseudocode}

\begin{document}
\title{Unsupervised Incremental Learning with Dual Concept Drift Detection for Identifying Anomalous Sequences\\
}


\author{
	\IEEEauthorblockN{
		Jin Li,
		Kleanthis Malialis,
        Christos G. Panayiotou,
		Marios M. Polycarpou
	}
 
	\IEEEauthorblockA{
		\textit{KIOS Research and Innovation Center of Excellence}\\
		\textit{Department of Electrical and Computer Engineering}\\
        University of Cyprus, Nicosia, Cyprus\\
		\{li.jin, malialis.kleanthis, christosp, mpolycar\}@ucy.ac.cy
		\\ORCID: \{0000-0002-3534-524X, 0000-0003-3432-7434, 0000-0002-6476-9025, 0000-0001-6495-9171\}
	}
}



\maketitle
\begin{abstract}
In the contemporary digital landscape, the continuous generation of extensive streaming data across diverse domains has become pervasive. Yet, a significant portion of this data remains unlabeled, posing a challenge in identifying infrequent events such as anomalies. This challenge is further amplified in non-stationary environments, where the performance of models can degrade over time due to concept drift. To address these challenges, this paper introduces a new method referred to as VAE4AS (Variational Autoencoder for Anomalous Sequences). VAE4AS integrates incremental learning with dual drift detection mechanisms, employing both a statistical test and a distance-based test. The anomaly detection is facilitated by a Variational Autoencoder. To gauge the effectiveness of VAE4AS, a comprehensive experimental study is conducted using real-world and synthetic datasets characterized by anomalous rates below 10\% and recurrent drift. The results show that the proposed method surpasses both robust baselines and state-of-the-art techniques, providing compelling evidence for their efficacy in effectively addressing some of the challenges associated with anomalous sequence detection in non-stationary streaming data.
\end{abstract}


\section{Introduction}

\IEEEPARstart {I}{n} recent years, there has been a notable surge in the abundance of streaming data across diverse application areas.

In numerous real-world scenarios, the generating process displays an intrinsic non-stationary phenomenon, referred to as concept drift. This phenomenon can be induced by diverse factors, such as seasonality or periodicity effects (e.g., water consumption in a drinking distribution network), and shifts in user interests/preferences or behavior (e.g., in recommendation systems). 

Anomaly detection is a vital tool across domains, crucial for pinpointing in real time deviations from normal data behavior. Yet, a key challenge lies in the dynamic nature of real-world systems. For instance, in drinking water networks, evolving water demand complicates the detection of hardware (sensor or actuator) faults or the detection of cyber-attacks, highlighting an ongoing challenge in adapting to shifting norms. 


The majority of anomaly detection systems commonly employ either signature-based methods or data mining-based methods, relying on labeled training data \cite{gomes2017survey}. However, acquiring labeled data in real-time applications can be expensive or, in some cases, impossible.

An anomalous sequence refers to a continuous anomalous pattern in data points within a continuous period of time \cite{chen2021deep}. For example, in water distribution networks, water leakage can last for days or even weeks, which can cause significant water and financial loss \cite{kyriakides2014intelligent}. Therefore, anomaly detection may be of critical importance in real-world applications. However, anomalous sequences might be wrongly classified as drift, resulting in deterioration of model performance. 

To tackle these challenges, we design and evaluate a new method, referred to as VAE4AS. Specifically, the key contributions of this work are the following:
\begin{enumerate}
	\item We develop VAE4AS (Variational AutoEncoder for identifying Anomaly Sequences), a VAE-based incremental learning algorithm with a dual concept drift detection mechanism. VAE4AS can detect both abrupt and recurrent concept drifts in the presence of anomalous sequences, and does not rely on supervision. One of its novel characteristics is its dual explicit concept drift detection module, which works in a synergistic manner with incremental learning for effective adaption to nonstationary environments. The operation of drift detection tests is in the latent space.

	\item We perform an empirical investigation employing both real-world and synthetic datasets to analyze VAE4AS. Additionally, through a comparative study, we illustrate that the proposed method consistently surpasses existing baseline and state-of-the-art techniques, underscoring its efficacy in overcoming some of the challenges outlined above.
\end{enumerate}

The paper is structured as follows: In Section~\ref{sec:background}, we present essential background material that is integral to comprehending the contributions of this paper. Section~\ref{sec:related} delves into the related work, offering a contextual overview. Our proposed method is discussed in Section~\ref{sec:method}, providing a detailed exploration of the approach. The experimental setup is described in Section~\ref{sec:exp_setup}. Section~\ref{sec:exp_results} provides an empirical analysis of the proposed approach, accompanied by comparative studies of various methods. Finally, Section~\ref{sec:conclusion} summarizes our findings and provides some concluding remarks.

\section{Background}\label{sec:background}
\textbf{Online learning} involves a continual data generation process, producing a set of examples at each time step $t$, denoted as $S = \{B^t\}_{t=1}^T$. Here, each batch is defined as $B^t = \{(x^t_i, y^t_i)\}^M_{i=1}$. The overall number of steps is denoted by $T \in [1, \infty)$, and the data is typically drawn from a long, potentially infinite sequence. The quantity of examples in each step is represented by $M$. In the case where $M=1$, this scenario is known as \textbf{one-by-one online} learning, while for $M > 1$, it is referred to as \textbf{batch-by-batch online} learning \cite{ditzler2015learning}. This study concentrates on one-by-one learning, denoted as $B^t = (x^t, y^t)$, which is especially pertinent for real-time monitoring. The examples are sampled from an unknown time-varying probability distribution $p^t(x, y)$, where $x^t \in \mathbb{R}^d$ represents a $d$-dimensional vector in the input space $X \subset \mathbb{R}^d$, $y^t \in \{1, ..., K\}$ denotes the class label, and $K \geq 2$ is the number of classes. In the context of this study, focusing on anomaly detection, the number of classes is set to $K=2$ ("normal" and "anomalous").

In the paradigm of one-by-one online classification, the model receives an individual instance $x^t$ at time $t$ and generates a prediction $\hat{y}^t$ based on a concept $h: X \to Y$. In the realm of \textbf{online supervised} learning, the model is provided with the true label $y^t$. Its performance is evaluated using a loss function, and subsequently, it undergoes training based on the incurred loss. This iterative process repeats at each time step. The continuous adaptation of the model without complete re-training, denoted as $h^t = h^{t-1}.train(\cdot)$, is referred to as \textbf{incremental} learning \cite{losing2018incremental}.

In real-time data streaming applications, promptly acquiring class labels is often impractical. To overcome this challenge, the research community has explored alternative learning paradigms such as \textbf{online semi-supervised} \cite{dyer2013compose} and \textbf{online active} \cite{zliobaite2013active} learning, as well as methods such as few-shot learning \cite{malialis2022nonstationary} and data augmentation \cite{malialis2022augmented}. While these have been demonstrated to be effective, they still rely on the availability of labeled data. In this study, our emphasis is on one-by-one \textbf{online unsupervised} learning, which operates without the need for any class labels, i.e., $B^t = (x^t)$.

Data nonstationarity is a prominent challenge observed in certain streaming applications, often attributed to concept drift, which refers to a change in the underlying joint probability distribution \cite{ditzler2015learning}. \textbf{Concept drift} can lead to shifts in the data characteristics over time. Specifically, the drift between two time steps $t_i$ and $t_j$, where $i \ne j$, is defined as follows:
\begin{equation}
\quad p^{t_i}(x,y) \neq p^{t_j}(x,y)
\end{equation}

\begin{figure}[h!]
\begin{subfigure}{.5\columnwidth}
  \centering
  \includegraphics[width=.89\columnwidth]{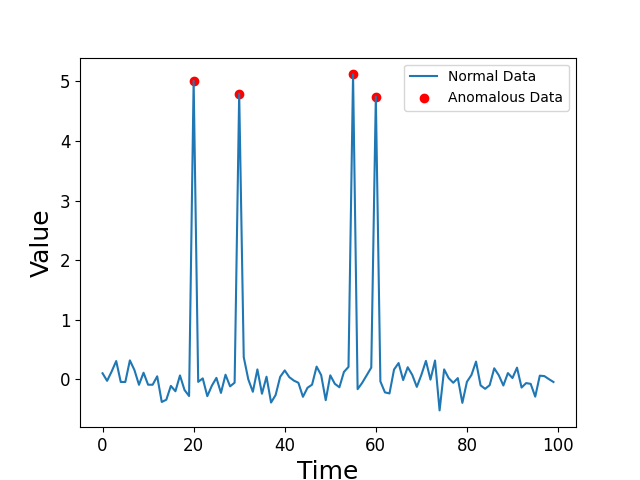}
  \caption{Anomalous points}
  \label{fig:anomalous_point}
\end{subfigure}%
\begin{subfigure}{.46\columnwidth}
  \centering
  \includegraphics[width=.97\columnwidth]{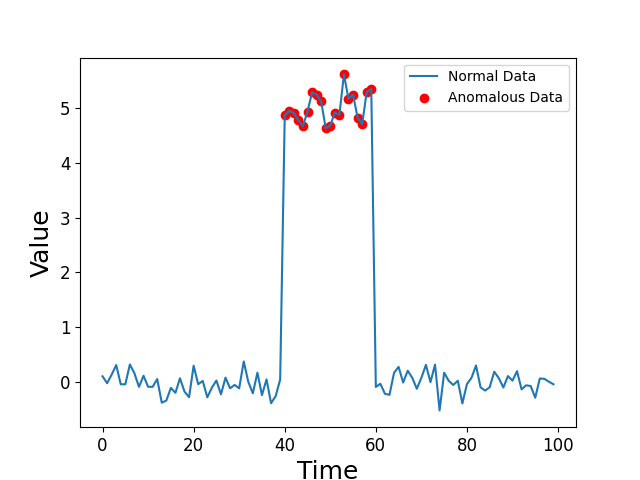}
  \caption{Anomalous sequence}
  \label{fig:anomalous_sequence}
\end{subfigure}
\caption{Illustration of anomalous points and sequences}
\label{fig:ap_as}
\end{figure}

\textbf{Anomalous sequence} refers to a set of continuously anomalous points. As illustrated in Fig.~\ref{fig:ap_as}, where $x$-axis represents time steps and $y$-axis represents the value of instances, a clear distinction can be observed between Fig.~\ref{fig:anomalous_point} and Fig.~\ref{fig:anomalous_sequence}. This study specifically centers on anomalous sequences. $t_{start}$ and $t_{end}$ represent the time steps when anomalous expression commences and concludes. Eq.~(\ref{eq:ano_expr}) provides the definition of an anomalous sequence. The anomalous sequence defined here can encompass both i.i.d. data and time-series data.

\begin{equation}\label{eq:ano_expr}
f(t) = \begin{cases} 
      \text{Normal Expression} & 0 \leq t < t_{\text{start}} \\
      \text{Anomalous Expression} & t_{\text{start}} \leq t < t_{\text{end}} \\
      \text{Normal Expression} & t_{\text{end}} \leq t 
\end{cases}
\end{equation}

Distinguishing between concept drift and anomalous sequences, even in supervised settings, remains a key research challenge. This work is concerned with developping an approach in unsupervised settings.

\section{Related Work}\label{sec:related}

\subsection{Concept drift adaptation}
Approaches addressing concept drift are commonly categorized as either passive or active\cite{ditzler2015learning}.

\subsubsection{Passive methods}
Passive strategies implicitly address drift using incremental learning. Within this classification, techniques can be further categorized into memory-based and ensemble methods. A memory-based algorithm typically employs a sliding window to retain a set of recent examples on which the model is trained. Representative methods include CVFDT \cite{hulten2001mining} and OS-ELM\cite{liang2006fast}. Ensembling involves utilizing a collection of models that can be dynamically added or removed based on their performance. Representative methods in this category include DDD \cite{minku2011ddd} and SEA\cite{street2001streaming}. Several research studies have been introduced to address imbalanced data in non-stationary environments, including AREBA \cite{malialis2020online} and ROSE\cite{cano2022rose}. 

\subsubsection{Active methods}
These strategies rely on explicitly identifying changes in the data distribution to initiate an adaptation mechanism \cite{ditzler2015learning}. Two primary categories of detection mechanisms have been investigated: statistical tests and threshold-based mechanisms. Statistical tests monitor the statistical characteristics of the generated data, while threshold-based mechanisms observe prediction errors and compare them against a predefined threshold. 

Autoencoders have also been employed as drift detectors. An autoencoder-based approach is presented by \cite{jaworski2020concept}, which focuses on detecting concept drift by monitoring two distinct cost functions: cross-entropy and reconstruction error. The variation in these cost functions serves as an indicator for concept drift detection.

\subsubsection{Other methods}
Hybrid approaches, such as HAREBA \cite{malialis2022hybrid}, have been proposed to combine the strengths of both active and passive methods. An alternative approach is called drift unlearning \cite{artelt2022unsupervised}, which attempts to revert the data distribution to the original one, prior to the concept drift.

\subsection{Online anomaly detection}
\subsubsection{Detection of anomalous points}

In anomaly detection, a typical approach involves training a classifier on normal data to establish a baseline of "normality", flagging any behavior that deviates from this baseline as anomalous. In contrast, the Isolation Forest (iForest) algorithm \cite{liu2008isolation} employs a fundamentally different strategy. It explicitly isolates anomalies instead of constructing normal profiles by building an ensemble of trees and identifying anomalies as instances with shorter average path lengths within the trees.

Recent advancements in anomaly detection have incorporated deep learning techniques, notably AEs. As outlined in \cite{chalapathy2019deep}, these methods offer substantial advantages due to their ability to learn hierarchical discriminative features from data, making them particularly effective in complex problem domains compared to traditional anomaly detection approaches. Several autoencoder-based methods have been proposed, including CPD\cite{mustafa2017unsupervised} and strAEm++DD\cite{li2023autoencoder}. CPD employs a deep abstract feature space stored in a sliding window, detecting change points through a \textit{log-likelihood ratio random walk} analysis on each point, triggering the training of a new model with examples from the sliding window. The anomaly detection mechanism of strAEm++DD involves computing the reconstruction loss for each instance within a moving window, ranking the losses, and setting a threshold at a certain percentile. Instances with reconstruction losses beyond this threshold are classified as anomalous.

In \cite{dong2018threaded}, a Streaming Autoencoder (SA) for online anomaly detection is introduced. SA utilizes an AE with incremental learning techniques to adapt to streaming data. Ensembling is employed, where the data stream is divided into threads, further segmented into buffer windows. This allows each ensemble member to be trained on a distinct part of the data stream.

Nevertheless, when confronted with an anomalous sequence, the aforementioned methods may fail to distinguish concept drift from the presence of anomalous sequences. This underscores the contribution of the proposed method.

\subsubsection{Detection of anomalous sequences}
Several algorithms are tailored to capture temporal relationships among instances. Among them, LSTM-VAE has emerged as a well-explored method in various studies, including \cite{niu2020lstm}, and \cite{fahrmann2022lightweight}, leveraging LSTM networks to effectively capture both long- and short-term dependencies inherent in sequential time-series data. 

While \cite{rosenberger2022extended} introduces a specific extension to address anomalous sequences, it is limited to one-dimensional time-series data. Another proposed method, DiFF-RF \cite{marteau2021random}, is designed to handle anomalous sequences but operates in a semi-supervised manner. In contrast to the methods mentioned above, the proposed approach VAE4AS aims to deal with situations of high-dimensional data in an unsupervised way.


\section{The VAE4AS Method}\label{sec:method}
In this section, we begin by providing an overview of the proposed method, VAE4AS. Following that, we delve into the anomaly detection component, including its incorporation into incremental learning for enhanced anomaly detection capabilities. Lastly, we elucidate the dual concept drift mechanism integrated within VAE4AS."

\begin{figure}[H]
	\centering
	\includegraphics[scale=0.65]{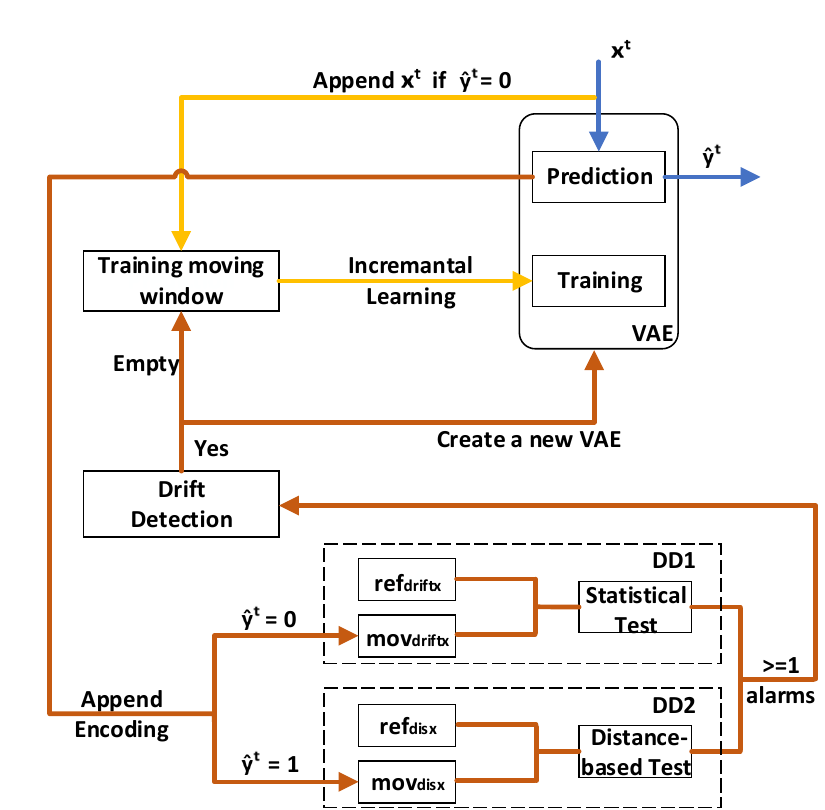}
	\caption{An overview of the VAE4AS design.}
	\label{fig:method}
\end{figure}

The overview of the proposed VAE4AS design is shown in Fig.~\ref{fig:method}. The prediction part is displayed in blue. The system first observes the instance $x^t \in \mathbb{R}^d$ at time $t$, and the VAE-based method outputs a prediction $y^t \in \{1, ..., K\}$. If the instance is classified as normal, it is then appended to the sliding windows $mov_{train}$ for incremental learning and its encoding is appended to $mov_{driftx}$ for statistical test. Otherwise, the encoding of classified anomalous instances is appended to the sliding windows $mov_{disx}$ for distance-based test.

The VAE is incrementally updated, using the data in memory, which is displayed in yellow color. Then we incorporate explicit concept drift detection as displayed in brown color. In the presence of drift, normal instances that have undergone drift may be classified as either normal or anomalous by the current classifier, depending on the nature of the drift. To address this, we introduce a dual drift detection (DD) methods DD1 and DD2, the details will be provided in a later section. Once an alarm flag is raised by any DD, the training window will be emptied and a new VAE will be created and trained. The corresponding pseudocode is shown in Algorithm 1.


\subsection{Anomaly detection and incremental learning}
\textbf{Model.} We consider a VAE\cite{kingma2013auto} which incorporates regularization to guide the encoder in learning a distribution $q(z|x)$ whihc is assumed to be a multivariate Gaussian. Assuming the encoding $z$ is k-dimensional ($z \in {R}^k$), the encoder produces two vectors of size $k$, representing the means $\mu \in {R}^k$ and standard deviations $\sigma \in \mathbb{R}^k$ for each dimension. The variance $\sigma^2$ corresponds to the diagonal elements of the covariance matrix. To generate samples from this distribution, the "reparameterization trick" is employed, where $z$ is generated as $z = \mu + \epsilon \odot \sigma$, with $\epsilon \sim N\left(0, I_k\right)$. In terms of regularization, the Kullback-Leibler (KL) divergence \cite{kullback1951information} is utilized. It measures the divergence between two probability distributions, specifically the KL divergence between the learned distribution with parameters $\mu$ and $\sigma$ for input $x$, and the unit Gaussian:

\begin{equation}\label{eq:kl}
\begin{aligned}
l_{K L}(x) & =K L\left(q(z \mid x) \| N\left(0, I_k\right)\right) \\
& =\frac{1}{2} \sum_{i=1}^k \mu_i^2+\sigma_i^2-\log \left(\sigma_i^2\right)-1
\end{aligned}
\end{equation}

The total loss is determined by the combination of the reconstruction loss and the regularization loss, represented by Eq.~(\ref{eq:vae}), in which $\beta \geq 0$ is a weight parameter that adjusts the importance of the KL divergence loss in the overall loss. $l_{A E}(x, \hat{x})$ denotes the loss of a traditional AE. For real-valued inputs, it is defined as the sum of squared differences, while for binary inputs it is defined as the cross-entropy:

\begin{equation}\label{eq:vae}
l_{V A E}(x, \hat{x})=l_{A E}(x, \hat{x})+ \beta * l_{K L}(x).
\end{equation}

\textbf{Memory.} The proposed method incorporates a dynamic mechanism using a sliding window, denoted as $mov_{train}$ with a size parameter $W_{train}$, to retain the most recent instances classified as normal. At any given time $t$, a queue $mov_{train}^t = \{x_t\}^t_{t-W_{train}+1}$ is maintained. Here, for any pair of instances $x_{i}$ and $x_{j}$ in $q^t$ with $j > i$, it holds that $x_{j}$ has been observed more recently in time than $x_{i}$. The sliding window can implicitly address the problem of concept drift, as obsolete examples will eventually drop out of the queue. However, it is acknowledged that due to the inherent imperfections in the classification process, the window may not be exclusively populated by normal instances. This point will be considered when anomaly detection is performed.

\textbf{Anomaly detection (prediction)}. The rationale behind employing a VAE lies in the expectation that the loss for an anomalous instance would substantially exceed that of a normal instance. This study adopts an adaptive threshold methodology, drawing inspiration from \cite{clark2018adaptive}. At each training time $t$, we calculate the loss of all the elements in the queue $mov_{train}$: $L^t=\{l(x^i,\hat{x}^i)\}^t_{i=t-W_{train}+1}$. The anomaly threshold at training time $t$ is set as follows:

\begin{equation}\label{eq:adtthreshold}
\theta^t = mean(L^t)+2*std(L^t).
\end{equation}

Establishing a threshold presents a challenge because delineating between normal data and outliers is often ambiguous. In this investigation, we introduce the adaptive threshold method, designed to autonomously grasp the statistical nuances of the data stream. As time advances, the threshold $\theta^t$ dynamically adjusts to align with the evolving patterns in the data stream.

After establishing a threshold, the process of anomaly prediction can be initiated. Let's consider a new instance $x^{t+\Delta}$, where $\Delta > 0$ signifies a point in time subsequent to the model's training at time $t$. If the cumulative loss associated with this instance surpasses the existing threshold $\theta^t$, the classification is made as anomalous:
\begin{equation}\label{eq:predict}
\hat{y}^{t+\Delta} =
\begin{cases}
1 \ \text{(anomaly)} & if \ l(x^{t+\Delta}, \hat{x}^{t+\Delta}) > \theta^t\\
0 \ \text{(normal)} & \text{otherwise}
\end{cases}
\end{equation}
\noindent where $t$ is the time of the most recent training, and $t+\Delta$ is the current time.

\textbf{Incremental learning (training)}. The cost function $J^t$ at training time $t$ is formulated as the mean loss computed across all instances within the queue:
\begin{equation}\label{eq:cost}
J^t = \frac{1}{W_{train}} \sum_{i=t-W_{train}+1}^t l(x^i,\hat{x}^i).
\end{equation}

\begin{algorithm}[h!]
\small
	\caption{VAE4AS}
	\label{alg:method}
	\begin{algorithmic}[2]
		
		\Statex \textbf{Input:} 
            \State $p$: window percentage replaced; $D$: unlabelled data for pre-training; $N$/$AN$: normal/anomalous data for reference drift/distance window; $W_{drift}$: window size for drift detection; $W_{distance}$: window size for Euclidean distance comparison; $W_{train}$: window size for re-training; $expiry\_time$: of the warning flag; $DIS_{thre}$: Euclidean distance threshold 

		\Statex \textbf{Init:}\Comment time $t=0$
        \State \textcolor{blue}{initialization of windows, flags and threshold}
		
		
  
	
        

		\Statex \textbf{Main:}
		\For{each time step $t \in [1, \infty)$}
		\State receive instance $x^t \in \mathbb{R}^d$
		\State predict $\hat{y}^t = h.predict(x^t) \in \{0,1\}$
      \If{ $\hat{y}^t == 0$ }
            \State append instance  $mov_{train}.append(x^t)$
            \State append instance $mov_{driftx}.append(x^t)$ 
            		\EndIf
     \If{ $\hat{y}^t == 1$ }
            \State append instance $mov_{AN}.append(x^t)$ 
            		\EndIf
       	\If{ ($mov_{train}.isFull()$ or $p\%$ replaced) and $flag_{warn} == False$}\Comment \textcolor{blue}{Incremental learning}
        		    \State $h.train(mov_{train})$
        	        \State $\theta = calc\_anomaly\_threshold(mov_{train})$
        		\EndIf
            
            \If{$ref_{driftx}.isFull()$}\Comment \textcolor{blue}{DD mechanism}
                \State $ref_{latent} = h.predict(ref_{driftx})$
  		    
            \Else
                \State append instance $ref_{driftx}.append(x^t)$
            \EndIf

    \If {$mov_{driftx}.isFull()$}
        \State $mov_{latent} = h.predict(mov_{driftx})$
    
        \If{$flag_{warn} == \text{False}$}
            \For{each dimension of $mov_{driftx}$ $i \in [1, \text{Dim}(mov_{driftx}))$}
                \State$p_{value}=KS(ref_{latent_i}, mov_{latent_i})$\Comment \textcolor{blue}{Eq.~(\ref{eq:ks}).}
                \If{ $p_{value}\leq P_{warn}$}
                    \State $flag_{warn} = \text{True}$ 
                \EndIf

            \If{$p_{value} \leq P_{alarm}$}
                \State $flag_{alarm} = \text{True}$ 
            \EndIf
            \EndFor
        \EndIf
    
    \EndIf

        \If {$flag_{warn}$ and $\neg flag_{alarm}$} \Comment \textcolor{blue}{DD1: Warning flag}
            \State append instance $mov_{warn}.append(x^t)$
            \If{flag raised for more than $expiry\_time$}
                \State $flag\_{warn} = False$
                \State $mov_{warn}$ empty
            \EndIf
        \EndIf

        \If {$flag_{alarm}$ or $Distance(ref_{disx}, mov_{AN}) > DIS_{thre}$} \Comment \textcolor{blue}{DD1, DD2: Alarm flag}
                \If {$DIS(ref_{disx}, mov_{AN}) > DIS_{thre}$}
                \State $h = train(mov_{AN})$                
                \Else
    		  \State $h = train(mov_{warn})$ 
                \EndIf

            \State reset windows, flags and threshold
        \EndIf

	    \EndFor
    \end{algorithmic}
\end{algorithm}

\noindent The autoencoder will be updated incrementally based on the cost incurred, that is, $h^t = h^{t-1}.train(J^t)$.

To avoid overfitting and to reduce the computational cost, training occurs when a specified percentage, denoted as $p\%$, of the sliding window is replaced. 

\subsection{Dual concept drift detection mechanism}
In this proposed method, we adopt a dual drift detection (DD) mechanism, one which is based on a statistical test (DD1) and the other which is distance-based (DD2).

\textbf{DD1: Statistical test}. As a non-parametric test, the Kolmogorov-Smirnov (KS) test does not rely on distribution assumptions. In this study, the KS test is adopted to compare the latent layer distribution of each dimension $i$. The KS test determines whether a significant difference exists by using the maximum difference of the cumulative distributions obtained by accumulating the two distributions as the statistical test quantity\cite{nitta2023detecting}. The $p_{value}$ is calculated as shown in Eq.~(\ref{eq:ks}), where $F(ref_{latent_i})$ and $F(mov_{latent_i})$ mean the cumulative distribution function of windows $ref_{latent_i}$ and $mov_{latent_i}$ respectively. $ref_{latent_i}=\{a_{j}\}_{j=1}^n$ and $mov_{latent_i}=\{b_{j}\}_{j=1}^n$, where $a_{j}, b_{j} \in \mathbb{R}$, correspond to the encoding as shown in Lines 15 and 19 in Algorithm 1. $N_{eff}$ is calculated by the size of $ref_{latent_i}$ and $mov_{latent_i}$, which is $W_{drift}$. $KS_{dis}$ represents the greatest distance between $F(ref_{latent_i})$ and $F(mov_{latent_i})$. $\gamma$ is calculated with $N_{eff}$ and $KS_{dis}$. 

\begin{equation}\label{eq:ks}
\begin{aligned}
p_{value}=2 \sum_{i=1}^{\infty}(-1)^{i-1} e^{-2 i^2 \gamma^2}\\
\text { where, } \gamma=\left(\sqrt{N_{eff}}+0.12+\frac{0.11}{\sqrt{N_{eff}}}\right) KS_{dis}\text {, }\\
KS_{dis}=\max \left|F(ref_{latent_i})-F(mov_{latent_i}) \right|\\
\text { and }
N_{eff}=\frac{W_{drift}^2}{2*W_{drift}}\\
\end{aligned}
\end{equation}

We will utilize this test to set two flags, $flag_{warn}$ and $flag_{alarm}$, as illustrated below. The warning flag signifies a cautionary signal for a potential concept drift, whereas the alarm flag is activated in the presence of an actual concept drift. The P-value for $flag_{warn}$ should be larger than the p-value for $flag_{alarm}$, i.e., $P_{warn} > P_{alarm}$. Once there is a flag alarm, then H0 (null hypothesis) is rejected and H1 (alternative hypothesis) is satisfied. The above description is shown in Lines 23-26 in Algorithm 1.
\begin{equation}\label{eq:pvalue}
{flag} =
\begin{cases}
warn \  & if \ P_{value} < P_{warn}\\
alarm \ &  if \ P_{value} < P_{alarm}\\
\end{cases}
\end{equation}

\textbf{DD2: Distance-based}. By calculating the Euclidean distance between the reference anomalous instances and the classified anomalous instances, we can identify the presence of drift as presented in Eq.~(\ref{eq:ed}) and Eq.~(\ref{eq:ed_flag}). The corresponding instances are stored in the windows $ref_{disx}$ and $mov_{AN}$. $ref_{disxij}$ and $mov_{ANij}$ represent the elements at the \(i\)-th row and \(j\)-th column of matrices $ref_{disx}$ and $mov_{AN}$ respectively. The elements of the row are the coordinates of a point along different dimensions and each column represents the values of a particular variable or feature across all observations. The threshold $DIS_{thre}$ is calculated offline. 

\begin{equation}\label{eq:ed}
\begin{aligned}
\text\ DIS(ref_{disx}, mov_{AN})=\sqrt{\sum_{i=1}^n \sum_{j=1}^m\left(ref_{disx i j}-mov_{AN i j}\right)^2}
\end{aligned}
\end{equation}

\begin{equation}\label{eq:ed_flag}
\begin{aligned}
{flag} = alarm \ &  if \ DIS(ref_{disx}, mov_{AN}) > DIS_{thre}\\
\end{aligned}
\end{equation}

\textbf{Warning flag raised}. The warning mechanism is only adopted for DD1. As shown in Line 27 of Algorithm 1, once a $flag_{warn}$ is raised and $flag_{alarm}$ is not raised, we start to store examples into $mov_{warn}$. In order to avoid false alarms, we set a parameter $expiry\_time$, if $flag_{warn}$ is raised for more than $expiry\_time$ and there is still no $flag_{alarm}$, we regard this as false warnings, and then reset the status of $flag_{warn}$ and empty $mov_{warn}$.

\textbf{Alarm flag raised}. When an alarm flag is triggered no matter by which detection mechanism, i.e., equal or more than one alarm is raised, a new autoencoder is instantiated to supersede the existing one. This new model undergoes training with the instances stored in $mov_{warn}$ or $mov_{AN}$. Simultaneously, the threshold is updated based on the contents of $mov_{warn}$. Following these updates, the windows, $mov_{train}$, $mov_{driftx}$, $mov_{warn}$, and $mov_{AN}$ are cleared, and the associated flags are reset. Subsequently, the new reference window post-drift $ref_{driftx}$ is replenished with arriving instances, with a size of $W_{drift}$, after the occurrence of drift.

\begin{table*}[h!]
\caption{Description of synthetic and real-world dataset}\label{tab:dataset}
\begin{adjustbox}{width=1.0\textwidth}
\begin{tabular}{|c|c|c|c|c|c|c|}
\hline
Dataset & \#Features & \#Arriving & Drift Time   & Anomalous Sequences                         & Before Drift                                                                                                                                          & After Drift                                                                                                                                            \\ \hline
Sea\cite{street2001streaming}     & 2          & 15000      & 5000, 10000  & (2000, 2100) (7000, 7100) (12000, 12100)   & \begin{tabular}[c]{@{}c@{}}Class 0: $x_1$+$x_2$\textgreater{}=10\\ Class 1: $x_1$+$x_2$\textless{}=3\\ $x_1$, $x_2$ in range {[}0,1{]}\end{tabular}                     & \begin{tabular}[c]{@{}c@{}}Class 0: $x_1$+$x_2$\textgreater{}=15\\ Class 1: $x_1$+$x_2$\textless{}=4\\ $x_1$, $x_2$ in range {[}0,1{]}\end{tabular}                      \\ \hline
Circle\cite{gama2004learning}  & 2          & 15000      & 5000, 10000  & (3000, 3200) (8000, 8200) (13000, 13200)   & \begin{tabular}[c]{@{}c@{}}Class 0: center=(0.6, 0.6), radius=0.2\\ Class 1: center=(0.2, 0.2), radius=0.2\\ $x_1$, $x_2$ in range {[}0,1{]}\end{tabular} & \begin{tabular}[c]{@{}c@{}}Class 0: center=(0.6, 0.6), radius=0.1\\ Class 1: center=(0.2, 0.2), radius=0.15\\ $x_1$, $x_2$ in range {[}0,1{]}\end{tabular} \\ \hline
Sine\cite{gama2004learning}    & 2          & 30000      & 10000, 20000 & (5000, 5050) (15000, 15050) (25000, 25050) & \begin{tabular}[c]{@{}c@{}}Class 0: $x_2$\textgreater{}sin($x_1$)+0.5\\ Class 1: $x_2$\textless{}sin($x_1$)-1\\ $x_1$ in {[}0, pi{]}, $x_2$ in {[}-1,1{]}\end{tabular}  & \begin{tabular}[c]{@{}c@{}}Class 0: $x_2$\textgreater{}sin($x_1$)\\ Class 1: $x_2$\textless{}sin($x_1$)-1.1\\ $x_1$ in {[}0, pi{]}, $x_2$ in {[}-1, 1{]}\end{tabular}    \\ \hline
Vib     & 10         & 22500      & 7500, 15000  & (3000, 3200) (9000, 9200) (17000, 17200)   & \begin{tabular}[c]{@{}c@{}}Class 0: mean=0, std=1\\ Class 1: mean=5, std=1\end{tabular}                                                               & \begin{tabular}[c]{@{}c@{}}Class 0: mean=3, std=1\\ Class 1: mean=0, std=0.5\end{tabular}                                                              \\ \hline
Fraud\cite{dal2015calibrating}   & 29         & 6100       & 3050         & (2000, 2200) (5050, 5250)                  & \cite{dal2015calibrating}                                                                                                            & \begin{tabular}[c]{@{}c@{}}Class 0: values of all features* 0.1\\ Class 1: values of all features*0.95\end{tabular}                                  \\ \hline
Wafer\cite{olszewski2001generalized}   & 152        & 5000       & 2500         & (1500, 1550) (4000, 4050)                  & \cite{olszewski2001generalized}                                                                                                      & \begin{tabular}[c]{@{}c@{}}Class 0: values of all features*0.5\\ Class 1: values of all features*0.95\end{tabular}                                  \\ \hline
\end{tabular}
\end{adjustbox}
\end{table*}

\begin{table}[h!]
\caption{Hyper-parameter values for VAE4AS}\label{tab:params_vae}
\begin{adjustbox}{width=0.5\textwidth}
\begin{tabular}{|c|cllllc|}
\hline
                                        & \multicolumn{1}{c|}{Sea} & \multicolumn{1}{c|}{Sine} & \multicolumn{1}{c|}{Circle} & \multicolumn{1}{c|}{Vib} & \multicolumn{1}{c|}{Fraud} & Wafer                    \\ \hline
Learning rate                           & \multicolumn{5}{c|}{0.001}                                                                                                                 & 0.0001                   \\ \hline
Mini-batch size                         & \multicolumn{6}{c|}{64}                                                                                                                                               \\ \hline
weight initializer                      & \multicolumn{6}{c|}{He Normal}                                                                                                                                        \\ \hline
Optimizer                               & \multicolumn{6}{c|}{Adam}                                                                                                                                             \\ \hline
Hidden activation                       & \multicolumn{6}{c|}{ Leaky ReLu}                                                                                                               \\ \hline
\multicolumn{1}{|l|}{Num. of epochs}    & \multicolumn{6}{c|}{10}                                                                                                                                               \\ \hline
\multicolumn{1}{|l|}{beta}              & \multicolumn{1}{l|}{1.0} & \multicolumn{1}{l|}{1.0}  & \multicolumn{1}{l|}{1.0}    & \multicolumn{1}{l|}{0.0} & \multicolumn{1}{l|}{1.0}   & \multicolumn{1}{l|}{1.0} \\ \hline
\multicolumn{1}{|l|}{Output activation} & \multicolumn{6}{c|}{Sigmoid}                                                                                                                                          \\ \hline
\multicolumn{1}{|l|}{Loss function}     & \multicolumn{4}{l|}{Binary cross-entropy}                                                                    & \multicolumn{2}{l|}{Square error}                     \\ \hline

\end{tabular}
\end{adjustbox}
\end{table}

\section{Experimental Setup}\label{sec:exp_setup}

\subsection{Datasets}
Our experimental study considers both synthetic and real-world datasets. For the sythetic datasets, we use the popular benchmark datasets Sine \cite{gama2004learning}, Sea\cite{street2001streaming} and Circle\cite{gama2004learning}. Furthermore, we have created Vib, a time-series dataset which consists of simulated equipment vibration data in industrial manufacturing. For the real-world datasets, we have used Fraud\cite{dal2015calibrating} which contains credit card transactions by European cardholders, and Wafer\cite{olszewski2001generalized}, a time-series dataset which relates to semi-conductor microelectronics fabrication. Additionally, both simple abrupt drift and recurrent drift are considered. Drift is not restricted to a single class, and can occur in either the normal or anomalous class. In all cases with concept drift, we have assumed that after its occurrence, no anomalous sequence follows for a reasonable amount of time (up to 1000 time steps). Lastly, the anomalous class rate varies from 0.5\% to 6.5\%. An overview of the datasets is provided in Table I, where the normal and anomalous classes are denoted with ``0'' and ``1'' respectively.

\subsection{Methods}

\textbf{Baseline}: In this approach, we initiate the pre-training process with unlabelled data. The training phase is conducted offline using 1800 normal examples as the training set, and the validation set comprises 200 normal instances and 50 anomalous instances. Pre-training is applied to all methods.

\textbf{strAEm++DD}: A state-of-the-art method for detecting anomalous points as described in Section III-B. It uses incremental learning and has a drift detection mechanism. For a fair comparison, we replace its AE with a VAE.

\textbf{iForest++}: An advanced tree-based method for detecting anomalous points as described in Section III-B. For fairness, incremental learning is adopted with the same window size as strAEm++DD.

\textbf{LSTM-VAE++}: A state-of-the-art method for detecting anomalous sequences as described in Section III-B. For fairness, incremental learning is adopted with the same window size as strAEm++DD.
   
\textbf{VAE4AS}: The proposed method as described in Section IV and its pseudocode is shown in Algorithm~\ref{alg:method}. For reproducibility, the hyper-parameters of VAE4AS are provided in Table II.

\subsection{Performance metrics}
One suitable and widely accepted metric that is insensitive to class imbalance is the geometric mean \cite{sun2006boosting}, defined as:
\begin{equation}\label{eq:gmean}
G\text{-}mean = \displaystyle \sqrt{R^+ \times R^-},
\end{equation}

\noindent where $R^+ = TP / P$ is the recall of the positive class, $R^-=TN / N$ is the recall (or specificity) of the negative class, and TP, P, TN, and N, are the number of true positives, total positives, true negatives, and total negatives, respectively.

\noindent Not only G-mean is insensitive to class imbalance, it has some important properties as it is high when all recalls are high and when their difference is small.

A widely used method for evaluating sequential learning algorithms is the prequential evaluation with fading factors, known for converging to the Bayes error in stationary data\cite{gama2013evaluating}. Its key advantage is dispensing with a holdout set, consistently testing the classifier on unseen data. We use a fading factor as 0.99 in all simulations, plotting the prequential metric (G-mean) in each time step averaged over 10 repetitions with error bars indicating the standard error.

\section{Experimental Results}\label{sec:exp_results}

\subsection{Empirical analysis of our method}
In this study, we set the window size $W_{drift}$ as 200 for real datasets and 1000 for synthetic datasets. The parameter $p$ is set to 100. In other words, the training occurs only after the whole sliding window is replaced. $W_{distance}$ is set to 50, $P_{warn}$ to 0.01, $expiry\_time$ to 100 and $P_{alarm}$ to 0.001.

\begin{figure}[H]
	\centering
	\includegraphics[scale=0.2]{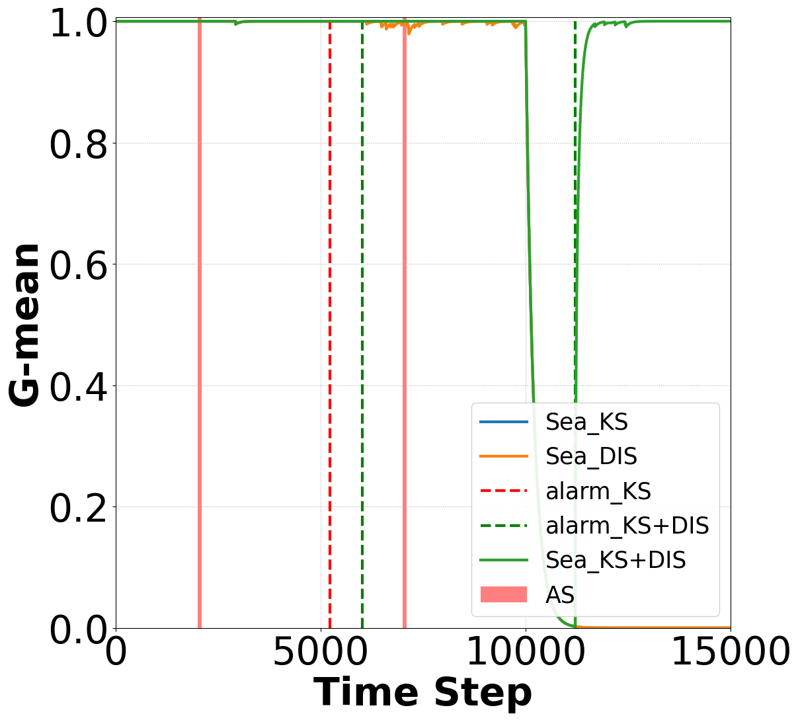}
	\caption{Performance of VAE4AS in non-stationary environments with different drift detection mechanisms.}
	\label{fig:dd_compare}
\end{figure}

\subsubsection{Drift Detection mechanism combination}
In this experiment, we demonstrate the importance of the combination of two drift detection mechanisms instead of relying solely on the KS test or distance measurement. The anomalous sequences (AS) are shown in red. As depicted in Fig.~\ref{fig:dd_compare}, when only KS test is employed, it can only detect the first drift, as indicated by the overlap of $alarm_{KS}$ and $alarm_{KS+DIS}$ at the left dashed line and the performance drops to zero after the second drift. In the case that only distance measurement mechanism is employed, no drift is detected and the performance drops to zero after the second drift as well. However, when dual drift detection mechanisms are employed together, two drifts are successfully detected. This is attributed to the fact that normal instances following the second drift are classified as anomalous, and they are not included in the window $mov_{driftx}$, preventing the KS test fails to being triggered.

\subsubsection{Role of the re-training window size}
\begin{figure}[h!]
\begin{subfigure}{.5\columnwidth}
  \centering
  \includegraphics[width=.9\columnwidth]{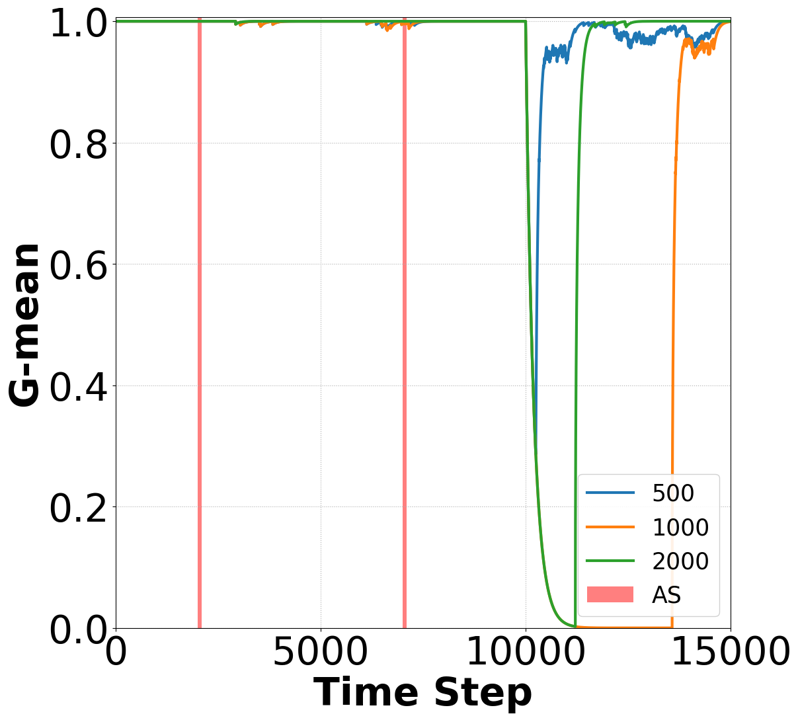}
  \caption{Sea}
  \label{fig:sea_wincompare}
\end{subfigure}%
\begin{subfigure}{.46\columnwidth}
  \centering
  \includegraphics[width=.98\columnwidth]{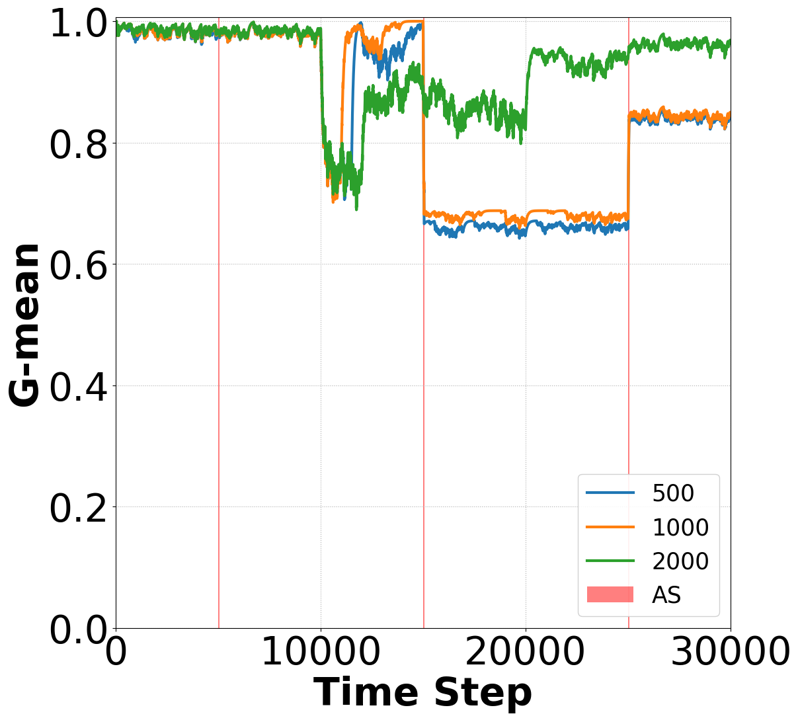}
  \caption{Sine}
  \label{fig:sine_wincompare}
\end{subfigure}
\caption{Performance of VAE4AS in non-stationary environments with different re-training window sizes.}
\label{fig:win_compare}
\end{figure}

In this experiment,we compare model performance and false alarm rates across various re-training window sizes $W_{train}$ using the Sea and Sine datasets. We set $expiry\_time=100$, $W_{distance}=50$, $P_{warn}=0.01$, and $P_{alarm}=0.001$. Table III summarizes the number of false alarms and performance metrics. For the Sea dataset, false alarms decrease from 5 to 0 as $W_{train}$ increases from 500 to 2000. Similarly, for the Sine dataset, the false alarm counts are 2, 1, and 1 for the models with $W_{train} = 500, 1000, 2000$ respectively. As illustrated in Fig.~\ref{fig:win_compare}, the model performance achieves the best or second-best results with a window size of 2000. Considering its impact on false alarms, we determine $W_{train} = 2000$ as the optimal setting, which will be adopted in subsequent experiments.

\subsection{Comparative study}
In this section, we compare five methods: Baseline, strAEm++DD, iForest++, and VAE4AS. Details and hyper-parameter values are in Section V-B and Table II. For synthetic datasets, $W_{train}$ is 2000; for real datasets, it is adjusted to 1000 for data volume.

The results are presented in Fig. 5. Notably, the performance of strAEm++DD significantly degrades upon encountering anomalous sequences, revealing its lack of robustness. iForest's performance varies considerably across different datasets, demonstrating its sensitivity to the dataset's nature. The performance of LSTM-VAE++ is quite unstable, especially after a drift or anomalous sequence, the performance of the model decreases significantly. While the Baseline exhibits impressive performance in certain scenarios (a), (d), and (f), it showcases a decline in performance for dataset Sine immediately after the first drift occurs. Moreover, in (d) and (e), the model's performance fluctuates substantially with the occurrence of anomalous sequences, and in (e), after the second anomalous sequence, the G-mean value drops to zero, underscoring the Baseline's sensitivity to anomalous sequences and drift.

\begin{figure}[!h]
  \centering
 \begin{subfigure}{.25\textwidth}
  \centering
  \includegraphics[width=0.9\columnwidth]{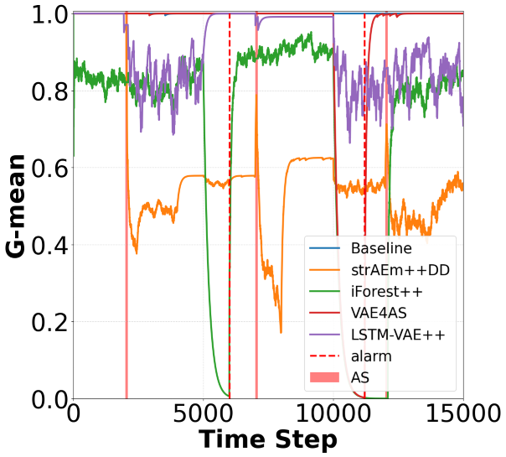} 
  \caption{Sea}
 \end{subfigure}%
 \begin{subfigure}{.25\textwidth}
  \centering
  \includegraphics[width=0.9\columnwidth]{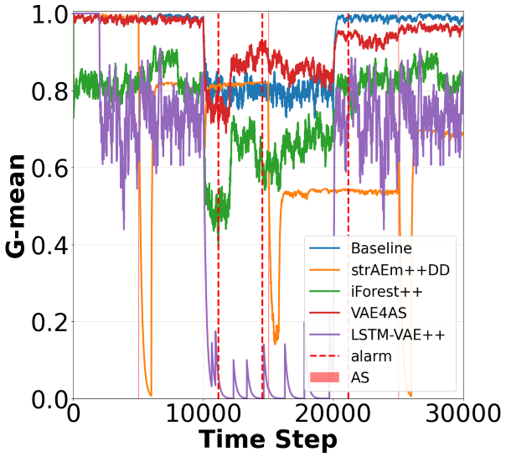} 
  \caption{Sine}
 \end{subfigure}%
 
 \begin{subfigure}{.25\textwidth}
  \centering
  \includegraphics[width=0.9\columnwidth]{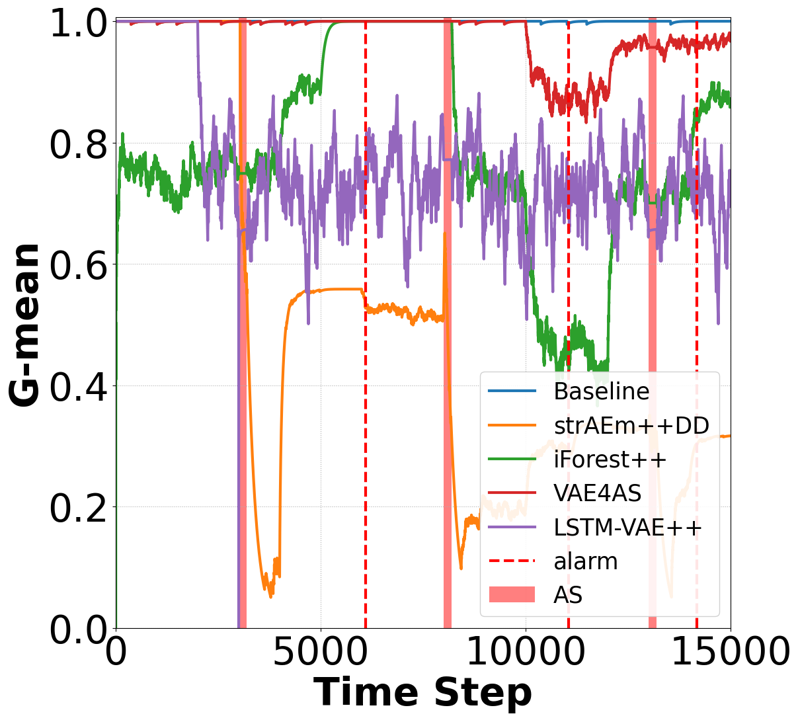} 
  \caption{Circle}
 \end{subfigure}%
  \begin{subfigure}{.25\textwidth}
  \centering
  \includegraphics[width=0.9\columnwidth]{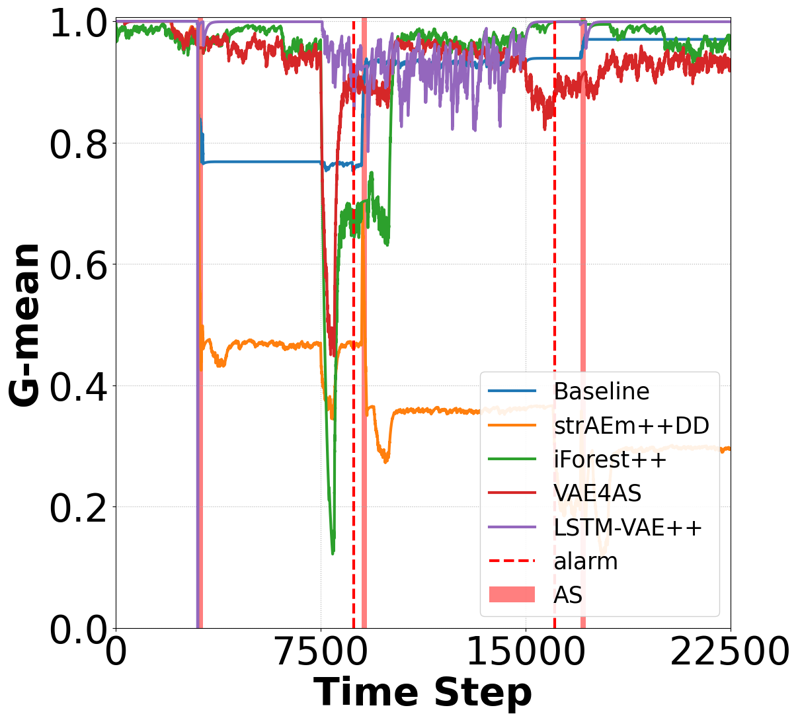} 
  \caption{Vib}
 \end{subfigure}%
 
  \begin{subfigure}{.25\textwidth}
  \centering
  \includegraphics[width=0.85\columnwidth]{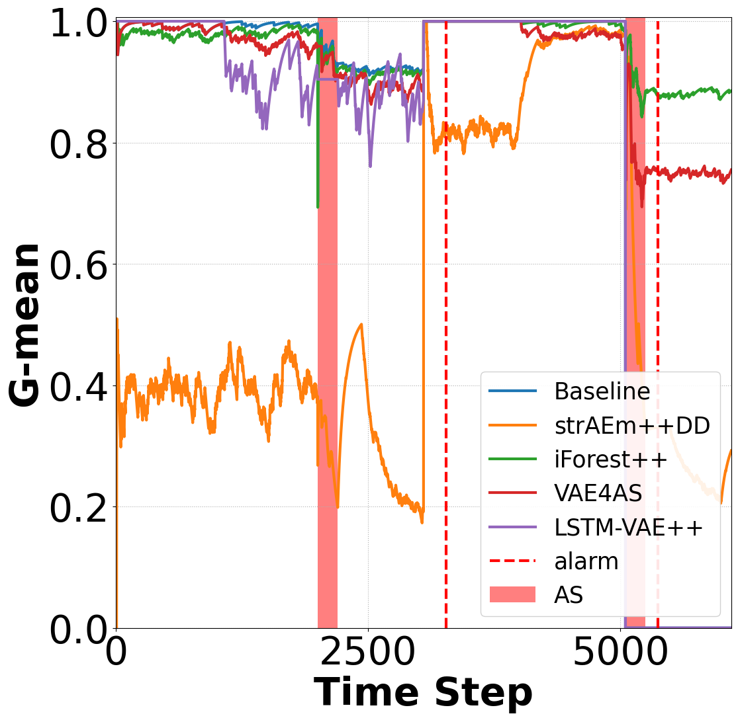} 
  \caption{Fraud}
 \end{subfigure}%
  \begin{subfigure}{.25\textwidth}
  \centering
  \includegraphics[width=0.9\columnwidth]{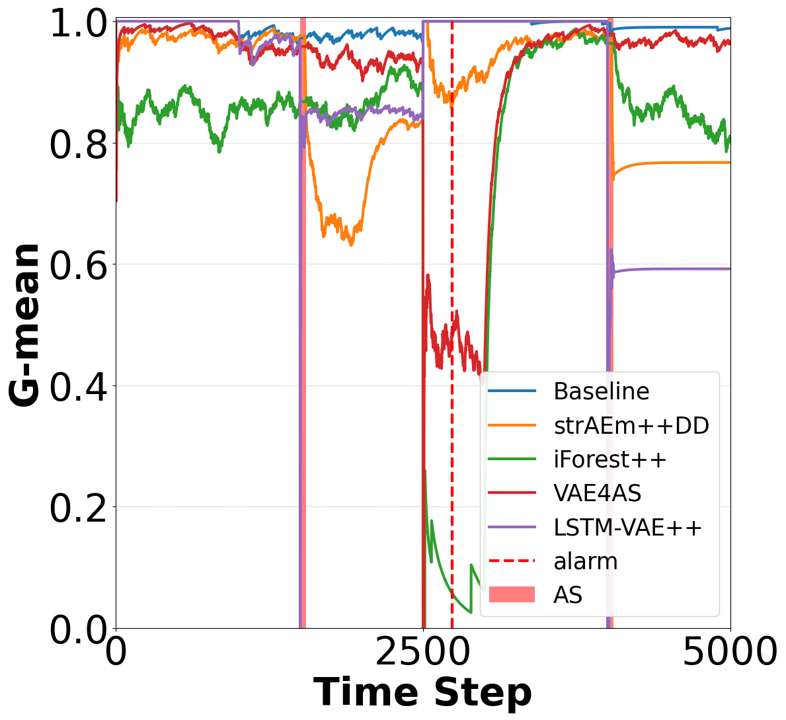} 
  \caption{Wafer}
 \end{subfigure}%
\caption{Comparison between Baseline, strAEm++DD, iForest++ and VAE4AS in nonstationary environments.}
\label{fig:compare}
\end{figure}

\begin{table}[h!]
\caption{Performance of VAE4AS and number of false alarms with different re-training window size}
\begin{tabular}{|c|ccc|ccc|}
\hline
Datasets              & \multicolumn{3}{c|}{Sea}                                     & \multicolumn{3}{c|}{Sine}                                    \\ \hline
Window size           & \multicolumn{1}{c|}{500}  & \multicolumn{1}{c|}{1000} & 2000 & \multicolumn{1}{c|}{500}  & \multicolumn{1}{c|}{1000} & 2000 \\ \hline
\#False alarms        & \multicolumn{1}{c|}{5}    & \multicolumn{1}{c|}{2}    & 0    & \multicolumn{1}{c|}{2}    & \multicolumn{1}{c|}{1}    & 1    \\ \hline
G-mean & \multicolumn{1}{c|}{0.98} & \multicolumn{1}{c|}{0.77} & 0.93 & \multicolumn{1}{c|}{0.84} & \multicolumn{1}{c|}{0.85} & 0.93 \\ \hline
\end{tabular}
\end{table}

In contrast, the proposed method, VAE4AS, consistently ranks among the top two algorithms, boasting a G-mean value exceeding 0.7 for all datasets. Moreover, even in the presence of recurrent drift, VAE4AS exhibits a reasonable number of false alarms, validating its robustness and effective detection capabilities. Overall, important remarks are as follows:

\begin{itemize}
    \item Performance-wise, VAE4AS significantly outperforms the strAEm++DD, iForest++, LSTM-VAE++ and Baseline methods.
    \item Furthermore, VAE4AS is more robust to different characteristics (i.e., drift type and anomalous rate) of the datasets as it appears to be less sensitive compared to the rest of the methods. The robustness of the proposed method can be attributed to the integration of a dual drift detection mechanism, effectively reducing the occurrence of false alarms and accurately identifying real drifts.
\end{itemize}

\section{Conclusion}\label{sec:conclusion}

Mining patterns from data streams presents formidable challenges, encompassing the unavailability of ground truth, the adaptation to nonstationary environments, and the presence of anomalous sequences. In such cases, distinguishing between concept drift and anomalous sequences constitutes a major challenge. In response to these challenges, we introduce a novel approach termed VAE4AS. This method leverages a variational autoencoder-based incremental learning strategy coupled with a dual drift detection mechanism. 
Our extensive experimental analysis establishes that the proposed VAE4AS method outperforms strong baseline and state-of-the-art methods. Looking ahead, our future research will explore the realm of multi-classification problems, specifically addressing scenarios involving the presence of multiple normal or anomalous classes. Additionally, we aim to apply our proposed method to more datasets with different concept drift characteristics to further validate the efficacy of our method.

\bibliographystyle{IEEEtran}
\bibliography{paper}

\end{document}